\documentclass[conference]{IEEEtran}
\IEEEoverridecommandlockouts
\usepackage{cite}
\usepackage{amsmath,amssymb,amsfonts}
\usepackage{algorithmic}
\usepackage{graphicx}
\usepackage{textcomp}
\usepackage{pgfplots}
\pgfplotsset{compat=1.18}
\usepackage{booktabs}
\usepackage{url}
\usepackage{balance}
\def\BibTeX{{\rm B\kern-.05em{\sc i\kern-.025em b}\kern-.08em
T\kern-.1667em\lower.7ex\hbox{E}\kern-.125emX}}

\usepackage{graphicx}
\usepackage{subcaption}  

\usepackage{amsthm}

\usepackage{makecell}

\newtheorem{theorem}{Theorem}

\theoremstyle{definition}

\newtheorem{remark}{Remark}

\begin{document}

\title{Characterizing Path-Independent Fees: A Route to Zero Impermanent Loss in CPMMs}

\author{
\IEEEauthorblockN{
  Andrey Voronin\IEEEauthorrefmark{1},
  Roman Vlasov\IEEEauthorrefmark{2},
  Vladimir Gorgadze\IEEEauthorrefmark{2}\IEEEauthorrefmark{3},
  Andrey Seoev\IEEEauthorrefmark{4}, 
  Yury Yanovich\IEEEauthorrefmark{5}
}
\IEEEauthorblockA{\IEEEauthorrefmark{1}%
  Novosibirsk State University, Novosibirsk, Russia
}
\IEEEauthorblockA{\IEEEauthorrefmark{2}%
  Moscow Institute of Physics and Technology, Moscow, Russia \\
}
\IEEEauthorblockA{\IEEEauthorrefmark{3}%
  IDEAS: Inter-Disciplinary \& Advanced Studies Center, Moscow, Russia \\
}
\IEEEauthorblockA{\IEEEauthorrefmark{4}%
  MEV-X, Moscow, Russia 
}
\IEEEauthorblockA{\IEEEauthorrefmark{5}%
  Skolkovo Institute of Science and Technology, Moscow, Russia \\
}
}

\maketitle

\begin{abstract}
Constant Product Market Makers use fees that are typically fixed proportions of trade size. When these fees are automatically reinvested into the pool, as in Uniswap~V2 and some designs of Uniswap V4, the final state after a trade can depend on how the trade is split into smaller transactions. This path dependence complicates the risk assessment for liquidity providers and affects composability guarantees. We characterize the functional class of fee structures that ensure path independence: the combined fee factor must depend only on the current pool invariant k=xy. For this class, we derive a system of ordinary differential equations governing pool dynamics and obtain a closed-form integral exchange formula. Within this class, we construct a parametric family of fee functions that achieve zero Impermanent Loss for a given initial pool state, and prove that no universal fee function can eliminate Impermanent Loss for all initial states simultaneously. We analyze implications for arbitrage windows and slippage, and validate our theory through controlled simulations. Our framework provides protocol designers with a principled approach to fee optimization that aligns liquidity provider and trader incentives while preserving composability.
\end{abstract}


\section{Introduction}
\label{sec:intro}

Decentralized exchanges (DEX) powered by Automated Market Makers (AMM) have become the foundational infrastructure in Decentralized Finance (DeFi). Among AMM designs, the Constant Product Market Maker (CPMM) remains one of the most widely deployed due to its mathematical simplicity, composability with other smart contracts, and robustness under adversarial conditions~\cite{tran2024financially}. Liquidity providers deposit pairs of tokens into these pools and earn fees from trades, but they face a well-known risk called Impermanent Loss (IL). IL arises when the relative price of the two tokens changes, causing the automated rebalancing mechanism to shift the portfolio composition. This loss is termed impermanent because it depends on the current price ratio and can diminish if prices revert, though accumulated fees may offset it.

Transaction fees serve as partial compensation for this risk. In most deployed systems, fees are set as fixed percentages of trade size and are automatically added back to the pool reserves~\cite{lebedeva2025dynamic}. This design choice has an important but often overlooked consequence: the final state of the pool after a trade may depend not only on the total volume exchanged but also on how that volume is split across multiple transactions. We call this phenomenon path dependence.

Path dependence matters for three practical reasons. First, it complicates risk modeling~\cite{melnikov2025smarter} for liquidity providers, who can no longer evaluate expected returns based solely on aggregate trade volume. Second, it creates opportunities for strategic behavior: arbitrageurs may fragment large trades to reduce effective fees or exploit slippage patterns. Third, it undermines composability guarantees that are central to DeFi~\cite{hossain2025crosslink}: smart contracts that interact with a pool may produce non-deterministic outcomes depending on transaction ordering and gas constraints.

Unlike order book exchanges where slippage depends on the depth of the limit order book at specific price levels, AMM slippage is deterministic and governed by the reserve ratio. However, when fees are state-dependent, this determinism breaks down across transaction fragments.

This paper addresses two related questions. First, what functional form of fees guarantees that the outcome of any swap depends only on the total volume, regardless of fragmentation? Second, within that class of fees, can we design mechanisms that eliminate IL for liquidity providers?

Our contributions are fourfold. We characterize the complete functional class of path-independent fee structures, showing that the combined fee factor must depend only on the current value of the pool invariant. We derive a system of ordinary differential equations that describes the pool dynamics under such fees and obtain a closed-form integral formula for the exchange outcome. Within this class, we construct a parametric family of fee functions that achieve zero IL for a given initial pool state, and we prove a negative result: no single fee function can eliminate IL for all possible initial states simultaneously. Finally, we analyze practical implications for arbitrage windows and slippage, and validate our theoretical findings through controlled numerical experiments.

The remainder of the paper is organized as follows. Section~\ref{sec:background} provides background on CPMMs and IL. Section~\ref{sec:related} reviews related work. Section~\ref{sec:problem} states the problem formally and presents our main results. Section~\ref{sec:proofs} contains the technical proofs. Section~\ref{sec:experiments} presents numerical experiments. Section~\ref{sec:discussion} discusses practical implications and limitations. Section~\ref{sec:conclusion} concludes the paper.

\section{Background: CPMM and Impermanent Loss}
\label{sec:background}

\subsection{Constant Product Market Makers}

A Constant Product Market Maker (CPMM) maintains a pool of two tokens, which we denote by reserves $x$ and $y$. The core invariant of the mechanism is that the product $k = x \cdot y$ remains constant in the absence of fees. When a trader wishes to exchange an amount $\Delta x$ of token $A$ for token $B$, the protocol computes the output amount $\Delta y$ such that the post-trade reserves satisfy the invariant. Solving $(x + \Delta x)(y - \Delta y) = xy$ yields the familiar formula $\Delta y = y \cdot \Delta x / (x + \Delta x)$.

The instantaneous exchange rate, or the marginal price, is given by the derivative $p = -dy/dx = y/x$. Because this price changes as trades execute, larger trades experience greater slippage: the effective price paid deviates more from the initial marginal price. This slippage is a fundamental feature of automated market makers and distinguishes them from order book exchanges, where slippage depends on the available liquidity depth at each price level rather than a continuous curve.

\subsection{Fees and Reinvestment}

In practice, AMMs charge fees on trades. For analytical completeness across different AMM designs, we consider a generalized fee structure with two parameters: $\gamma_1$ applied to the input token and $\gamma_2$ applied to the output token. Whilst fees are commonly applied to the input amount ($\gamma_1 < 1, \gamma_2 = 1$), some designs apply fees to the output or both sides. For example, protocols with two-sided fees exist, such as Curve. For Uniswap V2 and Balancer, the standard case simplifies to $\gamma_2=1$. 

When fees are automatically reinvested into the pool reserves, the invariant $k$ is no longer constant: it increases over time as fees accumulate. This reinvestment mechanism is used in Uniswap V2 and is an option in Uniswap V4 hook designs. The effective invariant change depends on the combined fee factor $\alpha = 1 - \gamma_1\gamma_2$~\cite{Vlasov2025Impact}.

\subsection{Impermanent Loss}

Impermanent Loss (IL) quantifies the difference between the value of a liquidity provider's position inside the pool and the value of holding the same initial amounts outside the pool, evaluated at current market prices. Crucially, for our analysis, the IL is calculated using the pool's marginal price after the trade, which arbitrageurs align with the external market price. Formally, if the prevailing price of the token $B$ in terms of the token $A$ is $p$, then the value of the holding portfolio is $V_{\text{hold}} = p \cdot x_0 + y_0$, while the pool position value is $V_{\text{pool}} = p \cdot x + y$, where $(x,y)$ are the current reserves. The absolute IL is $IL_a = V_{\text{pool}} - V_{\text{hold}}$.

When fees are present and reinvested, the analysis becomes more subtle: fees generate revenue that can offset losses from price movements. Recent work has introduced the concept of Impermanent Gain~\cite{Vlasov2025Impact}, referring to scenarios where fee income exceeds the loss from rebalancing~\cite{zeller2025automated,mohanty2025proactive}, resulting in a net positive outcome for the liquidity provider.

\section{Related Work}
\label{sec:related}

Automated Market Makers (AMMs) enable decentralized token exchange by replacing the traditional order book with a liquidity pool and a pricing function implemented in a smart contract~\cite{zetzsche2020decentralized,uniswapv2core2020}. The quality and efficiency of these decentralized markets compared to centralized exchanges have been empirically studied~\cite{Barbon2021}, revealing trade-offs in market depth and price discovery. In a two-asset AMM, the pool holds reserves $(x,y)$ and defines a trading rule via an invariant, e.g., a curve $F(x,y)=\text{const}$. The instantaneous exchange rate is given by the pool marginal price $p = -\,\mathrm{d}y/\mathrm{d}x$. Because this slope changes with the reserves, AMMs naturally exhibit slippage.

Impermanent Loss is classically defined as the difference between the value of a liquidity provider's position in an AMM and the value of the position holding the same assets outside the pool~\cite{Xu_2023, uniswapv2core2020, Aigner2021, Loesch2021}. Mathematical formalizations have refined this concept through the lens of "Greeks" for liquidity providers~\cite{Bardoscia2023} and generalized treatments across constant function market makers~\cite{Tangri2023}. When the relative market price changes, arbitrage trades rebalance the pool reserves, so the LP ends up with a different asset mix. This mix can be worth less than the original hold portfolio because of the AMM's convex rebalancing rule. Recent research has expanded this understanding by introducing the symmetrical concept of \textbf{impermanent gain}, reframing IL not merely as a loss, but as a transfer of value between counterparties. As shown by Kim et al.~\cite{Kim2022} the impermanent gain is the positive financial outcome for the AMM that arises from the exact same price movement that causes an impermanent loss for an individual LP. This duality is articulated by Labadie, who demonstrates that for AMMs with a constant-product formula, the trader's slippage is mathematically equivalent to the AMM's negative IL~\cite{Labadie2022}.

Popular AMM designs differ mainly in the choice of invariant. Uniswap V2 is based on the constant product invariant~\cite{uniswapv2core2020}, while Balancer generalizes this to a weighted constant product~\cite{martinelli2019balancer,cruz2025amm}. Comparative analyses show how different invariant curves lead to distinct IL landscapes~\cite{Kim2024}. Beyond basic invariant designs, concentrated liquidity markets such as Uniswap V3 have introduced more complex risk-return profiles~\cite{Cartea2023, Heimbach2022}.

A primary line of defense against net IL is the accumulation of swap fees. In~\cite{Vlasov2025Impact}, the authors study several AMM designs and emphasize that swap fees, when reinvested, make the pool invariant effectively non-constant over time, creating a non-trivial region of impermanent gain when trade size is sufficiently small. The strategic setting of fees~\cite{mancino2025decentralization,diallo2025optimized} is a complex optimization problem. Recent work explores \textbf{dynamic fee mechanisms} as a tool for IL mitigation. Baggiani et al.~\cite{baggiani2025optimaldynamicfeesautomated} formulate optimal dynamic fees as a stochastic control problem, revealing regimes to deter arbitrage or attract noise trading. However, most of this work treats fees as exogenous parameters and does not address the structural question of which fee functions preserve desirable properties, like path independence. To our knowledge, this paper provides the first characterization of path-independent fee structures for CPMMs.

\section{Problem Statement and Main Results}
\label{sec:problem}

\subsection{Formal Setup}

Consider a CPMM pool with reserves $(x,y)$ and invariant $k = xy$. Let $\gamma_1(x,y)$ and $\gamma_2(x,y)$ denote the fractions of input and output tokens, respectively, that participate in the invariant calculation after fees are applied. When a trader exchanges $\Delta x$ tokens of $A$ for $B$, the pool updates according to:
\begin{equation*}
\bigl(x + \gamma_1 \Delta x\bigr)\bigl(y - \Delta y\bigr) = k.
\label{eq:invariant_fee}
\end{equation*}

We say that a fee structure is \emph{path independent} if, for any total trade size $\Delta x$, the final pool state $(x_f, y_f, k_f)$ is the same regardless of whether the trade is executed as a single transaction or split into multiple smaller transactions.

\subsection{Main Result 1: Characterization of Path-Independent Fees}

\begin{theorem}[Path Independence Characterization]
\label{thm:pathind}
A fee structure is path independent if and only if there exists a function $\Phi: \mathbb{R}_+ \to (0,1]$ such that
\begin{equation}
1 - \gamma_1(x,y)\gamma_2(x,y) = \Phi(xy) = \Phi(k).
\label{eq:phi_condition}
\end{equation}
\end{theorem}

This result reduces the design space for path-independent fees to a single-variable function of the invariant $k$. Intuitively, the combined effect of input and output fees must depend only on the current "size" of the pool, not on the individual reserve levels.

\begin{remark}[On fee decomposition]
Theorem~\ref{thm:pathind} constrains only the combined fee factor 
$\alpha = 1 - \gamma_1\gamma_2$. Individual components $\gamma_1(x,y)$ and 
$\gamma_2(x,y)$ may vary freely provided their product satisfies 
$\gamma_1\gamma_2 = 1 - \Phi(xy)$. This flexibility allows protocol designers 
to allocate fee incidence between input and output tokens (e.g., input-only 
fees as in Uniswap V2, or symmetric splits) without affecting path independence.
\end{remark}

\subsection{Main Result 2: Integral Exchange Formula}

For any path-independent fee $\Phi$, the pool dynamics during a trade can be described by a system of ordinary differential equations. Parameterizing the trade by cumulative input $s \in [0, \Delta x]$, we have:
\begin{align}
\frac{dx}{ds} &= 1, \label{eq:dx} \\
\frac{dy}{ds} &= -\frac{(1-\Phi(k))\,y}{x}, \label{eq:dy} \\
\frac{dk}{ds} &= \Phi(k)\,\frac{k}{x}. \label{eq:dk}
\end{align}

\begin{theorem}[Integral Exchange Formula]
\label{thm:exchange}
Let $G(k) = \int \frac{dk}{\Phi(k)\,k}$. Then for any path-independent fee $\Phi$,
\begin{equation*}
G(k_f) - G(k_0) = \ln\!\left(1 + \frac{\Delta x}{x_0}\right).
\label{eq:exchange_formula}
\end{equation*}
The final reserves are $x_f = x_0 + \Delta x$ and $y_f = k_f / x_f$.
\end{theorem}

This closed-form expression allows protocol designers to compute the outcome of any trade without simulating the full trajectory, provided the fee function $\Phi$ is known.

\subsection{Main Result 3: Zero Impermanent Loss Construction}

\begin{theorem}[Zero-IL Fee for Fixed Initial State]
\label{thm:zeroil}
For a fixed initial invariant $k_0$, there exists a fee function $\Phi_{k_0}(k)$ such that the absolute IL is zero for all trade sizes $\Delta x \geq 0$. This function is given parametrically by:
\begin{equation}
\Phi_{k_0}\!\left(k_0 \cdot \frac{(1+\alpha)^2}{1+2\alpha}\right) = \frac{2\alpha}{1+2\alpha}, \qquad \alpha = \frac{\Delta x}{x_0}.
\label{eq:phi_zeroil}
\end{equation}
\end{theorem}

The zero-IL fee satisfies $\Phi_{k_0}(k_0) = 0$, meaning infinitesimal trades 
at the reference state incur no fee. However, for any finite trade size 
$\alpha > 0$, the fee becomes positive as $k$ moves away from $k_0$. 
The invariant grows according to the integral of fees along the trajectory 
(Theorem~\ref{thm:exchange}), and this accumulated fee revenue exactly 
compensates for the impermanent loss. This is why $IL_a = 0$ holds for all 
$\alpha \geq 0$, not just locally at $k_0$.

\subsection{Main Result 4: Impossibility of Universal Zero-IL Fees}

\begin{theorem}[No Universal Zero-IL Fee]
\label{thm:impossibility}
There exists no function $\Phi(k)$ such that the absolute IL is zero for all initial states $k_0 > 0$ and all trade sizes $\Delta x > 0$ simultaneously.
\end{theorem}

This negative result has important practical implications: zero-IL fees must be state-aware. Protocols can either recompute the fee function when liquidity is added or removed, or target zero loss only for a reference state.

\section{Technical Proofs}
\label{sec:proofs}

\subsection{Proof of Theorem~\ref{thm:pathind}}

Path independence requires that the differential change in the invariant, $dk = (1-\gamma_1\gamma_2)(y\,dx + x\,dy)$, be an exact differential. Writing $\alpha(x,y) = 1 - \gamma_1\gamma_2$, the integrability condition for the form $\alpha y\,dx + \alpha x\,dy$ to be exact is $\partial_x(\alpha y) = \partial_y(\alpha x)$. This simplifies to the partial differential equation $x \partial_x \alpha = y \partial_y \alpha$.

The method of characteristics shows that the solutions to this PDE are constant along the curves where $xy = \text{const}$. Therefore, $\alpha(x,y) = \Phi(xy)$ for some function $\Phi$, which establishes the necessity of condition~\eqref{eq:phi_condition}. Sufficiency follows by direct substitution: if $\alpha = \Phi(xy)$, then $dk = \Phi(k)(y\,dx + x\,dy) = \Phi(k)\,d(xy)$ is exact because $d(xy) = y\,dx + x\,dy$.

\subsection{Proof of Theorem~\ref{thm:exchange}}

From equation~\eqref{eq:dx}, we have $x(s) = x_0 + s$. Substituting into~\eqref{eq:dk} gives:
\[
\frac{dk}{ds} = \frac{\Phi(k) \cdot k}{x_0 + s}.
\]
Separating the variables yields $\frac{dk}{\Phi(k) k} = \frac{ds}{x_0 + s}$. Integrating from $s=0$ to $s=\Delta x$:
\begin{eqnarray*}
\int_{k_0}^{k_f} \frac{dk}{\Phi(k) k} = \int_0^{\Delta x} \frac{ds}{x_0 + s} \\ = \ln(x_0 + \Delta x) - \ln(x_0) = \ln\!\left(1 + \frac{\Delta x}{x_0}\right).
\end{eqnarray*}
By the definition of $G$, the left-hand side equals $G(k_f) - G(k_0)$, completing the proof.

\subsection{Proof of Theorem~\ref{thm:zeroil}}

Setting $IL_a = 0$ in the general expression for IL yields the target trajectory $k_f(\alpha) = k_0 (1+\alpha)^2 / (1+2\alpha)$. Differentiating the identity $G(k_f(\alpha)) - G(k_0) = \ln(1+\alpha)$ with respect to $\alpha$ gives:
\[
G'(k_f) \frac{dk_f}{d\alpha} = \frac{1}{1+\alpha}.
\]
Using $G'(k) = 1/(\Phi(k)k)$ and substituting $k_f(\alpha)$ and its derivative yields a differential equation for $\Phi$. Solving this equation and expressing the result in parametric form yields~\eqref{eq:phi_zeroil}.

\subsection{Proof of Theorem~\ref{thm:impossibility}}
Assume for contradiction that a universal zero-IL fee function $\Phi(k)$ exists.
Fix an arbitrary target invariant $k^* > 0$. For any initial state $k_0 < k^*$, 
Theorem~\ref{thm:zeroil} implies that achieving $IL_a = 0$ when trading from 
$k_0$ to $k^*$ requires:
\begin{equation}
\Phi(k^*) = \frac{2\alpha(k_0)}{1 + 2\alpha(k_0)}, 
\quad \text{where} \quad 
k^* = k_0 \cdot \frac{(1+\alpha(k_0))^2}{1+2\alpha(k_0)}.
\label{eq:phi_constraint}
\end{equation}
The mapping $k_0 \mapsto \alpha(k_0)$ defined implicitly by 
\eqref{eq:phi_constraint} is strictly decreasing (verified by differentiation). 
Since $f(\alpha) = 2\alpha/(1+2\alpha)$ is strictly increasing in $\alpha$, 
the composition $k_0 \mapsto f(\alpha(k_0))$ is strictly decreasing. 
Therefore, distinct initial states $k_0^{(1)} \neq k_0^{(2)}$ yield distinct 
required values $\Phi(k^*)$, contradicting the assumption that $\Phi$ is a 
well-defined function. $\square$

\section{Numerical Experiments}
\label{sec:experiments}

We validate our theoretical results through controlled numerical experiments. All simulations use initial reserves $x_0 = y_0 = 100$, total trade size $\Delta x = 10$ (10\% of reserves), and fee parameters calibrated to typical Uniswap V2 settings ($\phi = 0.003$ for constant fees). The reference invariant is $k_0 = x_0 y_0 = 10^4$. For path-independent fee functions $\Phi(k)$, we compute the final invariant $k_f$ using the integral formula from Theorem~\ref{thm:exchange}; for standard Uniswap V2 implementations, we use the discrete update rule with fee reinvestment after each sub-swap.

\subsection{Visualizing Path Independence}

To build geometric intuition for Theorem~\ref{thm:pathind}, we visualize the fee factor $\alpha(x,y) = 1 - \gamma_1\gamma_2$ over the reserve space. For path-independent fees, the level sets of $\alpha$ should be parallel to the invariant hyperbolas $xy = \text{const}$, since $\alpha = \Phi(xy)$ depends only on the product.

\begin{figure*}[t]
\centering
\begin{subfigure}[b]{0.45\textwidth}
\centering
\includegraphics[width=\linewidth]{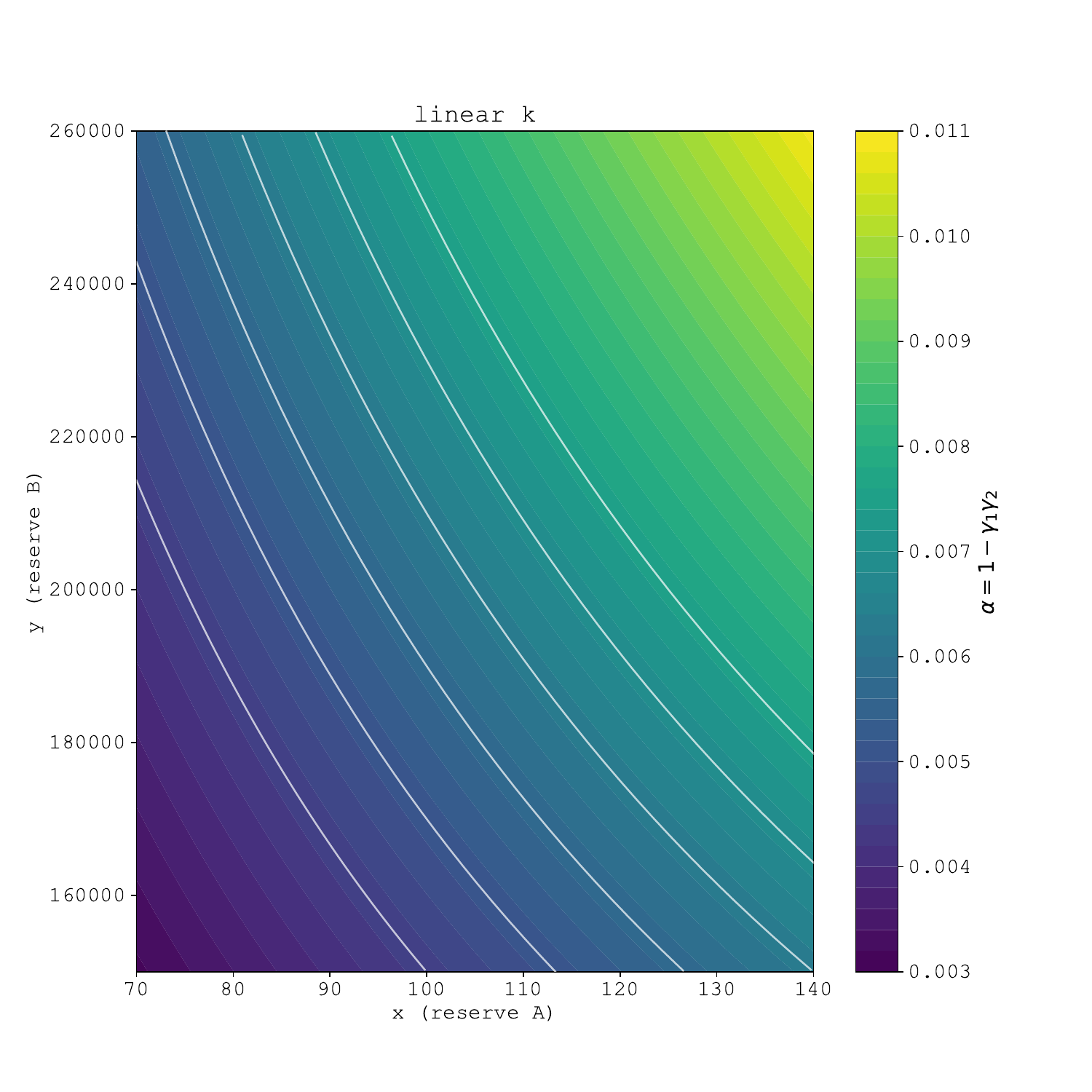}
\caption{Path-independent fee $\Phi(xy)$: contours align with invariant hyperbolas $xy=\text{const}$.}
\label{fig:linearK}
\end{subfigure}
\hfill
\begin{subfigure}[b]{0.45\textwidth}
\centering
\includegraphics[width=\linewidth]{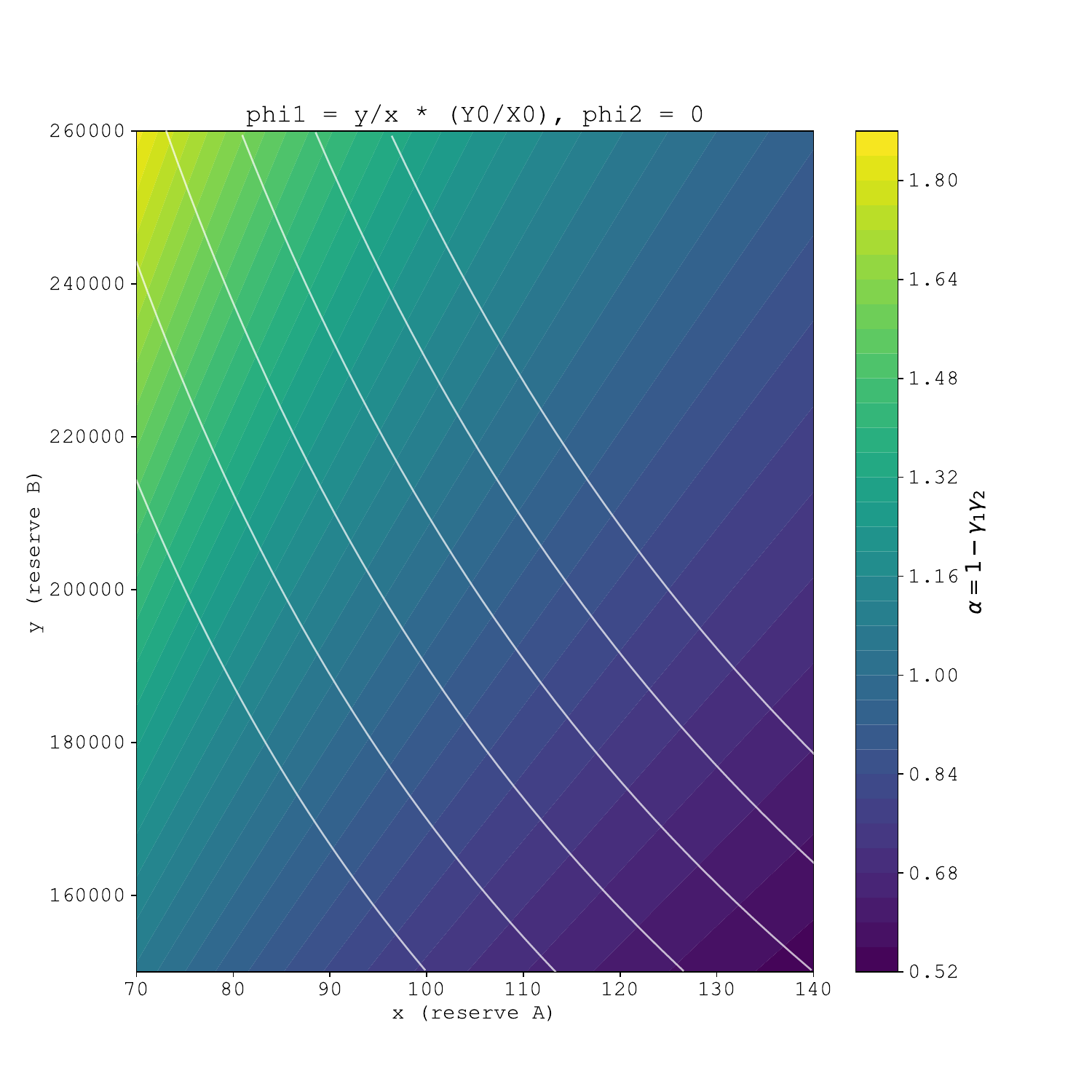}
\caption{Path-dependent fee $\Phi(y/x)$: contours cross hyperbolas, causing path dependence.}
\label{fig:proportional}
\end{subfigure}
\caption{Visualization of fee factor $\alpha(x,y) = 1-\gamma_1\gamma_2$ over the reserve space. Path-independent fees (left) have level sets parallel to characteristics $xy=\text{const}$, ensuring the exchange outcome depends only on total volume. Path-dependent fees (right) vary along characteristics, making the final state sensitive to trade fragmentation.}
\label{fig:characteristics}
\end{figure*}

Figure~\ref{fig:characteristics} compares two cases. In the path-independent case (left), where $\Phi = \Phi(xy)$, the color contours perfectly align with the hyperbolas, confirming that $\alpha$ is constant along the characteristics. In the path-dependent case (right), where $\alpha$ depends on the price ratio $y/x$, the contours cross the hyperbolas diagonally. This visualization explains why path dependence arises: even when moving along an invariant curve (constant $xy$), the fee rate changes if $\alpha$ depends on $x$ or $y$ separately, causing the final state to depend on the fragmentation of the trade.

\subsection{Splitting Test: Validating Theorem~\ref{thm:pathind}}

We implement a splitting test to quantitatively validate Theorem~\ref{thm:pathind}. For a fixed total trade size $\Delta x$, we compute the final invariant $k_f(N)$ when the trade is split into $N$ equal sub-trades. The relative error
\begin{equation}
\text{Error}(N) = \frac{|k_f(N) - k_f(1)|}{k_f(1)}
\label{eq:rel_error}
\end{equation}
measures deviation from the atomic-trade baseline $k_f(1)$. For path-independent fees, the theory predicts $\text{Error}(N) \approx 0$ for all $N$; for path-dependent fees, we expect persistent deviation that may grow with $N$.

\begin{figure}[t]
\centering
\includegraphics[width=0.9\columnwidth]{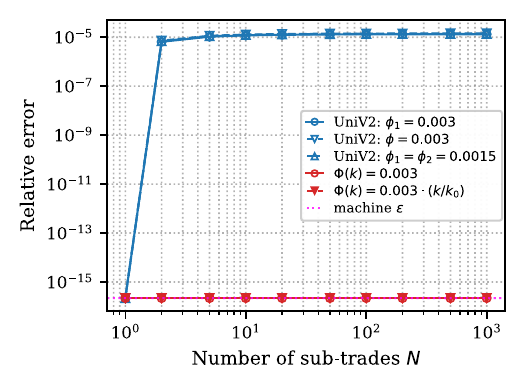}
\caption{Relative error $|k_f(N) - k_f(1)|/k_f(1)$ vs.\ number of sub-trades $N$ (log-log scale), with $x_0=y_0=100$, $\Delta x=10$. Path-independent $\Phi(k)$ designs (red) converge to machine precision ($\varepsilon$, magenta), validating Theorem~\ref{thm:pathind}. Standard Uniswap V2 implementations (blue) exhibit small errors ($\sim 10^{-5}$) due to discrete fee compounding; this deviation scales as $O(\Phi^2\alpha^2/N)$ and is negligible for typical fees ($\Phi \approx 0.003$) and trade sizes ($\alpha \lesssim 0.1$).}
\label{fig:splitting_test}
\end{figure}

Figure~\ref{fig:splitting_test} validates Theorem~\ref{thm:pathind}: path-independent $\Phi(k)$ designs (red) achieve machine precision ($\varepsilon \approx 2.2 \times 10^{-16}$) for all $N \geq 1$, confirming exact path independence. Uniswap V2 standard implementations (blue) exhibit small errors ($\sim 10^{-5}$) due to discrete fee compounding. This deviation scales as $O(\Phi^2\alpha^2/N)$ where $\Phi=0.003$ and $\alpha=0.1$; for these modest parameters the deviation is negligible in practice, but confirms that exact path independence requires the continuous integral formulation. As $N \to \infty$, all curves converge, recovering the continuous limit.

\subsection{Relative Effective Price: User Perspective}

We evaluate the effective price experienced by traders under different fee designs. The relative effective price is defined as:
\begin{equation}
p_{\text{rel}} = \frac{p_{\text{eff}}}{p_{\text{no-fee}}} = \frac{(\Delta y / \Delta x)_{\text{with fee}}}{(\Delta y / \Delta x)_{\text{no fee}}},
\label{eq:rel_price}
\end{equation}
where $p_{\text{eff}}$ is the actual exchange rate received by the user and $p_{\text{no-fee}}$ is the ideal rate without fees. Values $< 1$ indicate price degradation due to fees.

\begin{figure}[t]
\centering
\includegraphics[width=0.9\columnwidth]{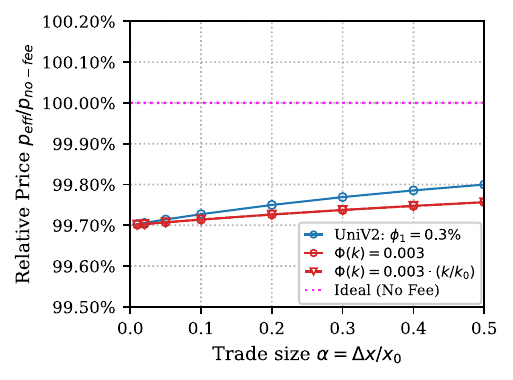}
\caption{Relative effective price $p_{\text{eff}} / p_{\text{no-fee}}$ vs.\ trade size $\alpha = \Delta x/x_0$. Standard Uniswap V2 with input fee $\phi_1=0.3\%$ (blue) and path-independent designs with $\Phi(k) = 0.003$ (red solid) both yield relative prices close to the ideal baseline (magenta). The linear $\Phi(k) = 0.003 \cdot (k/k_0)$ design (red dashed) shows slightly lower relative prices for larger trades, reflecting its adaptive fee behavior.}
\label{fig:relative_price}
\end{figure}

Figure~\ref{fig:relative_price} shows that all fee designs yield relative prices within $0.2\%$ of the ideal baseline for small trades ($\alpha \lesssim 0.1$), reflecting the modest fee magnitude ($\phi = 0.003$) and the limited invariant drift ($\Delta k/k_0 \approx \Phi\alpha \approx 3\times 10^{-4}$). The two red curves $\Phi(k)$ overlap closely because this small drift is insufficient to trigger a noticeable adaptation in the linear design $\Phi(k) = 0.003\cdot(k/k_0)$; a significant divergence would require larger trades or cumulative fee accumulation. The UniV2 curve (blue) lies marginally closer to the ideal baseline than the symmetric $\Phi(k)$ implementations due to its input-only fee incidence, which concentrates the fee impact differently than the balanced $\gamma_1=\gamma_2$ split used for the evaluation $\Phi(k)$. For larger trades ($\alpha \gtrsim 0.3$), the linear $\Phi(k)$ design exhibits slightly lower relative prices, demonstrating its adaptive mechanism: as $k$ grows during the trade, $\Phi(k)$ increases, extracting more value to offset impermanent loss at the cost of higher effective fees for large trades.

\subsection{Zero-IL Fee Performance}

We evaluate the zero-IL fee construction from Theorem~\ref{thm:zeroil}. For a reference state $k_0$, the fee function $\Phi_{k_0}(k)$ is designed to achieve $IL_a = 0$ for all trade sizes $\alpha = \Delta x / x_0$ starting from $k_0$.

\begin{figure}[t]
\centering
\includegraphics[width=0.9\columnwidth]{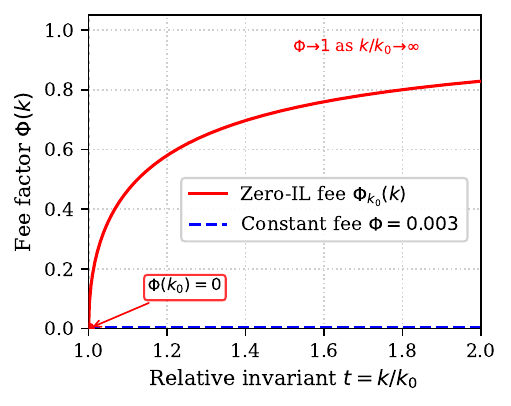}
\caption{Zero-IL fee function example $\Phi_{k_0}(k)$ (red) vs.\ constant fee example $\Phi = 0.003$ (blue dashed). 
The state-aware fee satisfies $\Phi_{k_0}(k_0) = 0$ and increases monotonically for $k > k_0$, 
charging higher rates for larger deviations to offset impermanent loss.}
\label{fig:zeroil_fee_shape}
\end{figure}

Figure~\ref{fig:zeroil_fee_shape} plots $\Phi_{k_0}(k)$ against the relative invariant $t = k/k_0$. The red curve shows the parametric fee function, while the blue dashed line represents a constant fee baseline ($\Phi = 0.003$, i.e., 0.3\%). The plot confirms two key properties: (i) $\Phi_{k_0}(k_0) = 0$, meaning infinitesimal trades at the reference state incur no fee; (ii) $\Phi_{k_0}(k)$ increases monotonically for $k > k_0$, charging higher rates for larger deviations to offset impermanent loss.

While the zero-IL fee function $\Phi_{k_0}(k)$ theoretically approaches~1 as $k/k_0 \to \infty$, 
practical implementations only require its behavior in a narrow neighborhood of the reference state. 

Key observations:
\begin{itemize}
    \item The zero-IL fee starts at $\Phi(k_0) = 0$ and grows as 
    $\Phi_{k_0}(k) \approx 2\sqrt{k/k_0 - 1}$ for small deviations $k \approx k_0$.
    \item For a typical constant fee of 0.3\% ($\Phi = 0.003$), the zero-IL fee remains 
    \emph{lower} than the constant fee only when $k/k_0 \lesssim 1.000002$ 
    (i.e., invariant deviation $\lesssim 0.0002\%$), and exceeds it thereafter.
    \item This implies that the zero-IL design charges \emph{less} than a fixed 0.3\% fee 
    only for extremely small trades near the reference state, while charging \emph{more} 
    for virtually all practical trade sizes to offset impermanent loss.
\end{itemize}
This steep adaptive behavior--near-zero fees for infinitesimal trades, rapidly increasing 
fees for any finite deviation--is precisely what enables $IL_a = 0$ while maintaining 
reasonable trader costs only in the limit of vanishing trade size.

These results confirm that zero-IL fees are inherently state-aware: they provide exact protection only at the reference state $k_0$ for which they were constructed. Practical deployment would require either (i) periodic recomputation of $\Phi_{k_0}$ when liquidity is added/removed, or (ii) acceptance of residual IL when the pool drifts significantly from $k_0$.

\section{Discussion}
\label{sec:discussion}

\paragraph{Key insights}
Our experiments validate three core findings: (i)~only fee functions with $\Phi = \Phi(k)$ achieve exact path independence (Theorem~\ref{thm:pathind}), with Uniswap V2 showing negligible $O(10^{-5})$ discretization errors for typical parameters; (ii)~zero-IL fees $\Phi_{k_0}(k)$ are inherently state-aware, providing exact protection only near their reference state $k_0$; (iii)~Theorem~\ref{thm:impossibility} establishes that universal zero-IL fees cannot exist, forcing designers to choose between state-aware updates, constant fees with residual IL, or adaptive approximations.

\paragraph{Practical implications}
Path-independent fees preserve composability and admit closed-form arbitrage/slippage analysis. On-chain implementation of $\Phi_{k_0}(k)$ requires solving a quadratic per trade (feasible in Ethereum gas limits); for efficiency, the protocols can precompute lookup tables with piecewise-linear interpolation. Zero-IL fees induce slightly higher slippage for large trades--a necessary trade-off for LP protection.

\paragraph{Limitations and future work}
Our analysis covers two-asset CPMMs; extending to concentrated liquidity (Uniswap V3) or multi-asset pools requires handling position-specific invariants. Future work could explore data-driven adaptive fees, decentralized governance for updating $\Phi_{k_0}$, or hybrid designs blending constant and adaptive components.

\section{Conclusion}
\label{sec:conclusion}

We have characterized the functional class of path-independent fee structures for CPMMs, derived closed-form exchange formulas, and constructed a parametric family of fee functions that eliminate impermanent loss for a reference pool state. Our impossibility result clarifies fundamental limits: universal zero-IL fees cannot exist, motivating adaptive or state-aware designs.

These results provide protocol designers with a principled foundation for fee optimization. Path-independent fees preserve composability guarantees and simplify risk modeling; zero-IL constructions offer a blueprint for protocols targeting specific liquidity ranges; and the inherent trade-offs guide practical implementation choices.

By bridging structural analysis with experimental validation and implementation considerations, this work advances the design of sustainable and predictable liquidity provision mechanisms in DeFi.

\bibliographystyle{IEEEtran}
\bibliography{references}

\end{document}